\documentclass[letter]{aa}

\usepackage{graphicx}%
\usepackage{multirow}%
\usepackage{epstopdf}%
\usepackage{txfonts}
\usepackage{amsmath,amssymb,amsfonts}%
\usepackage{mathrsfs}%
\usepackage[title]{appendix}%
\usepackage{xcolor}%
\usepackage{textcomp}%
\usepackage{manyfoot}%
\usepackage{booktabs}%
\usepackage{algorithm}%
\usepackage{algorithmicx}%
\usepackage{algpseudocode}%
\usepackage{listings}%
\usepackage{comment}%
\usepackage{subcaption}%
\usepackage{overpic}%
\usepackage{multirow}%
\usepackage{hyperref}%

\begin{document}

\title{Fusion of JWST data: Demonstrating practical feasibility}

\author{Landry Marquis\inst{1}\inst{2}\inst{4}, Claire Guilloteau\inst{3}, Thomas Oberlin\inst{2}, Nicolas Dobigeon\inst{4} \and Olivier Berné\inst{1}}
 \institute{Institut de Recherche en Astrophysique et Planétologie, Université de Toulouse, CNRS, CNES, 31028 Toulouse, France
 \and Institut Supérieur de l'Aéronautique et de l'Espace ISAE-SUPAERO, Université de Toulouse, 31055 Toulouse, France
 \and Laboratoire d'Informatique Signal et Image de la Côte d'Opale, Université du Littoral Côte d’Opale, 62228 Calais, France
 \and Institut de Recherche en Informatique de Toulouse, Université de Toulouse, CNRS, Toulouse INP, 31400 Toulouse, France }
\date{Accepted February 23, 2026}

\abstract{
Data fusion is a computational process widely used in Earth observation to generate high-resolution hyperspectral data cubes with two spatial and one spectral dimensions. It merges data from instruments with complementary characteristics: one with low spatial but high spectral resolutions, and another with high spatial but low spectral resolutions. In astronomy, the use of such instrumental combinations is becoming increasingly common, making data fusion a promising approach for enhancing observational data. Until now, however, its application to astronomical data has remained unsuccessful. We present the first successful astronomical data fusion using JWST integral field spectroscopy with NIRSpec and imaging across 29 filters with NIRCam. Applied to observations of the d203-506 protoplanetary disk in Orion and of Titan, our method produces fused hyperspectral cubes with NIRCam spatial and NIRSpec spectral resolutions. These results pave the way for extracting the physical properties from JWST data with unprecedented spatial resolution and showcase the transformative potential of data fusion in astronomy.}


\titlerunning{Fusion of JWST data}
\authorrunning{L. Marquis, C. Guilloteau, T. Oberlin, N. Dobigeon \and O. Berné}
\maketitle
\nolinenumbers

\section{Introduction}

\begin{figure}[h!]
\centering
\includegraphics[width=0.5\textwidth]{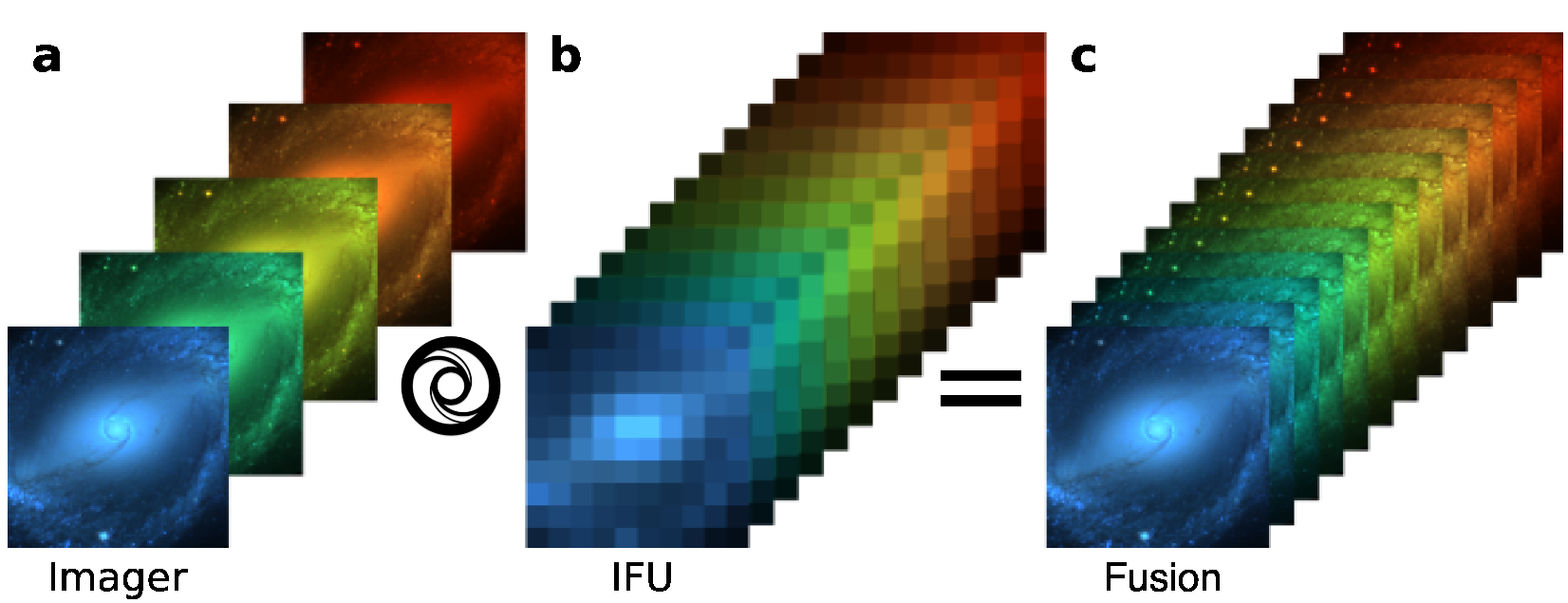}
\caption{Schematic illustration of astronomical data fusion. A high spatial resolution multispectral image (\textbf{a}) is fused with a high spatial resolution hyperspectral image (\textbf{b}) to produce a high spatial and spectral resolution hyperspectral cube (\textbf{c}). Credits: ESA/Hubble, NASA}\label{fig:principe_fusion}
\end{figure}

Combining multiple datasets obtained with different instruments is a common practice in astrophysics. Usually, different datasets are used independently to obtain complementary information on the chemical or physical characteristics of astronomical sources. Alternatively, one can directly combine some parts of datasets. For instance, \citet{bacon2023muse} separated sources in MUSE data cubes by incorporating Hubble Space Telescope data. A similar but certainly more ambitious goal consists of performing a comprehensive data fusion in which complementary observations are merged into a single data cube that retains the highest spatial and spectral resolutions of each set (Fig.~\ref{fig:principe_fusion}). While such a data fusion process has been widely used in Earth observation \citep{yokoya2017hyperspectral}, fusion of astronomical data has not yet been accomplished due to several main challenges, notably that the astronomical wavelength ranges are typically large enough to result in non-negligible spectral variations of the optical point spread function (PSF) \citep{soulez2013restoration, hadj2017restoration}, which significantly increases the complexity of the fusion process.

Despite these challenges, some recent works have already investigated the fusion of synthetic astronomical data. Their main achievements include designing computationally efficient algorithms \citep{guilloteau2020hyperspectral, pineau2023exact, pineau2025multispectral, lascar2025best} and generating realistic simulated data \citep{guilloteau2020simulated}. However, none of the previous studies has considered real astronomical data. In this work, we demonstrate that—by leveraging the latest in situ instrument models, the comprehensive Near Infrared Camera (NIRCam) and Near Infrared Spectrograph (NIRSpec) documentation, and rigorous cross-calibration—a practical fusion of JWST data is now achievable.

\section{Method}

\begin{figure*}[t]
    \includegraphics[width=0.65\linewidth]{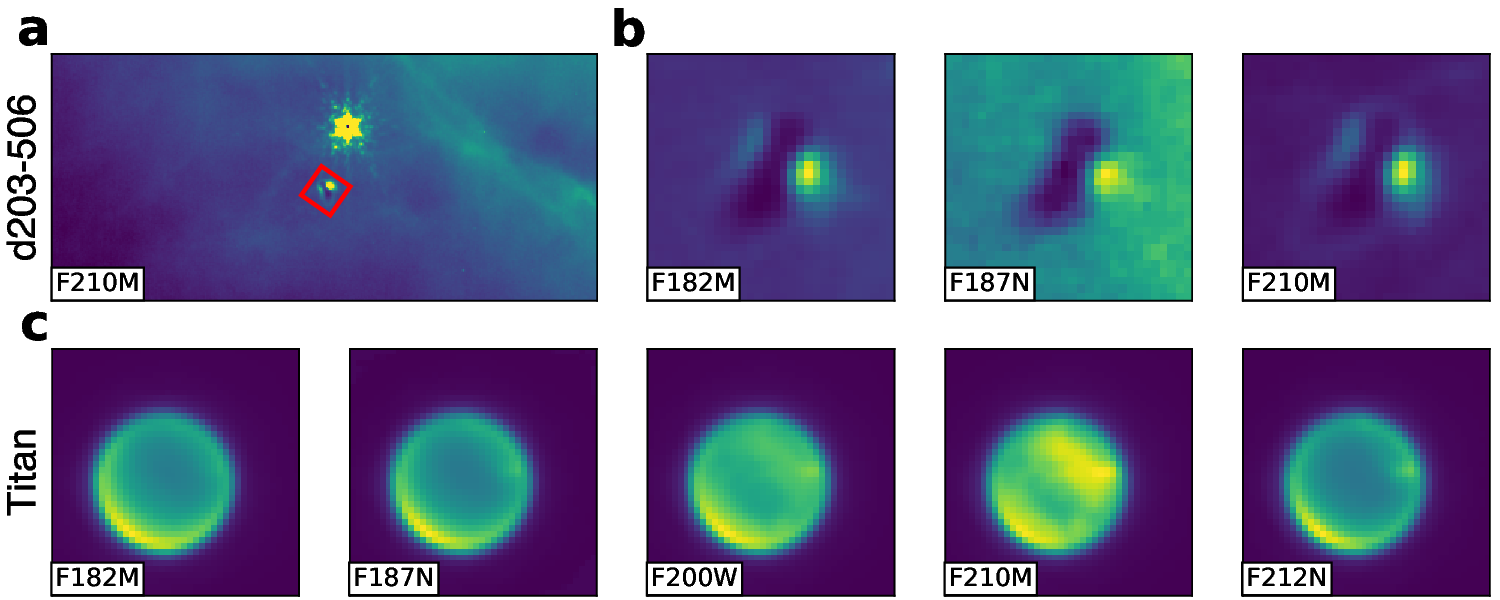}
\centering
\caption{Selected JWST NIRCam images. Large field of view NIRCam image (\textbf{a}) of the Orion Bar in the F210M filter. The red square denotes the field of view selected for the d203-506 protoplanetary disk. 1 x 1'' images (\textbf{b}) of the d203-506 protoplanetary disk observed by NIRCam filters F182M, F187N and F210M. 1.4 x 1.4'' images \textbf{c} of Titan observed by NIRCam filters F182M, F187N, F200W, F210M, and F212N.}
\label{fig:NIRCam}
\end{figure*}

The JWST conducts astronomical observations in the near- to mid-infrared range with unprecedented quality \citep{gardner2023james}. Among its instruments, the NIRCam \citep{rieke2023NIRCamthroughput} imager and the NIRSpec integral field unit (IFU) \citep{boker2022NIRSpecIFU} covers a spectral range from 0.6 to 5~$\mathrm{\mu m}$. The NIRCam imager is composed of 29 filters and two acquisition channels: the short wavelength channel, below 2.35~$\mathrm{\mu m}$, which has a 0.031~arcsec pixel scale, and the long wavelength channel, above 2.35~$\mathrm{\mu m}$, which has a 0.063~arcsec pixel scale \citep{STScINIRCamoverview}. The NIRSpec IFU has a 0.1~arcsec pixel scale and four high resolution filter and disperser combinations \citep{STScINIRSpecIFUoverview}. Its resolving power is around 2700, thus providing more than 9600 spectral channels. 

We denote by $Y_{\mathrm{m}}$ the multispectral image acquired by the NIRCam imager and by $Y_{\mathrm{h}}$ the hyperspectral cube acquired by the NIRSpec IFU over the same scene. From this pair of data, we aim to reconstruct the underlying fused hyperspectral cube $X$ defined at the NIRCam spatial resolution and the NIRSpec spectral resolution. To solve this problem, we framed the fusion task as a regularised inverse problem that first requires modelling of the observational processes. The high spatial and spectral resolution cube ($X$) to be reconstructed is assumed to be related to the NIRCam observations ($Y_{\mathrm{m}}$) according to the forward model
\begin{equation}
Y_{\mathrm{m}} \approx \textsf{NIRCam}(X),
\label{eq:NIRCam}
\end{equation}
where $\textsf{NIRCam}(\cdot)$ stands for the operator accounting for the effects of the NIRCam imager throughput and PSF. Similarly for the NIRSpec IFU, the forward model is written as%
\begin{equation}
Y_{\mathrm{h}} \approx \textsf{NIRSpec}(X),
\label{eq:NIRSpec}
\end{equation}
where $\textsf{NIRSpec}(\cdot)$ resumes the spatial and spectral degradations resulting from the NIRSpec throughput, PSF, and spatial sub-sampling. It is worth mentioning that for the two instruments the PSF varies heavily across the wavelengths. The instrument models are detailed in Appendix \ref{sec:prepro}. Regarding the throughputs, the results reported hereafter were obtained by exploiting the latest available in situ measurements \citep{rieke2023NIRCamthroughput,giardino2022NIRSpecthroughput}. We resorted to the theoretical PSF models derived by the \texttt{WebbPSF} simulation tool \citep{perrin2014WebbPSF} for their balance of simplicity and accuracy over newer alternatives \citep{nie2023hybpsf}. Equations \eqref{eq:NIRCam} and \eqref{eq:NIRSpec} relate the observation models to the observations through approximations. The expected errors underlying the approximation symbols ($\approx$) are related to imperfect instruments models, data pre-processing, and external and instrumental noises.

At this point, data fusion boils down to jointly inverting the two instrument models to recover the fused image, $X$. However, despite its linearity, this inversion is ill-posed, making solutions non-unique and highly sensitive to model mismatches and noise. The strategy we propose consists of reformulating the fusion task as a regularised least squares problem:
\begin{equation} 
    \min_X \|Y_{\mathrm{m}} - \textsf{NIRCam}(X)\|^2 + \gamma \|Y_{\mathrm{h}} - \textsf{NIRSpec}(X)\|^2 + R(X),
\label{eq:fusion}
\end{equation}
where $\gamma$ is a parameter adjusting the importance of the NIRSpec data fidelity term and $R(\cdot)$ is a regularisation. This formulation of the fusion task departs from the alternative that would exploit the noise statistics to derive a penalised log-likelihood. Indeed such an alternative requires establishing a proper noise model, which is very challenging without ensuring substantial enhancements of the fusion results \citep{pontoppidan2016pandeia, guilloteau2020simulated,guilloteau2020hyperspectral}. 
Through regularisation, we promote expected spectral and spatial properties exhibited by the fused cube $X$. First, $X$ is assumed to be linearly represented by a few elementary spectra. Hence, the adopted spectral regularisation imposes a low-rank structure on the data cube $X$, which is constrained to belong to an affine subspace of lower dimension. This spectral subspace is spanned by the most relevant spectra identified beforehand by a principal component analysis of the NIRSpec data. Then, the spatial regularisation is derived from a Sobolev norm to promote a smooth spatial content of the fused image. This choice has the advantage of leading to a globally quadratic minimisation problem, which can be easily solved (see Appendix \ref{sec:fusion}).

\begin{figure*}[t]
    \includegraphics[width=0.65\linewidth]{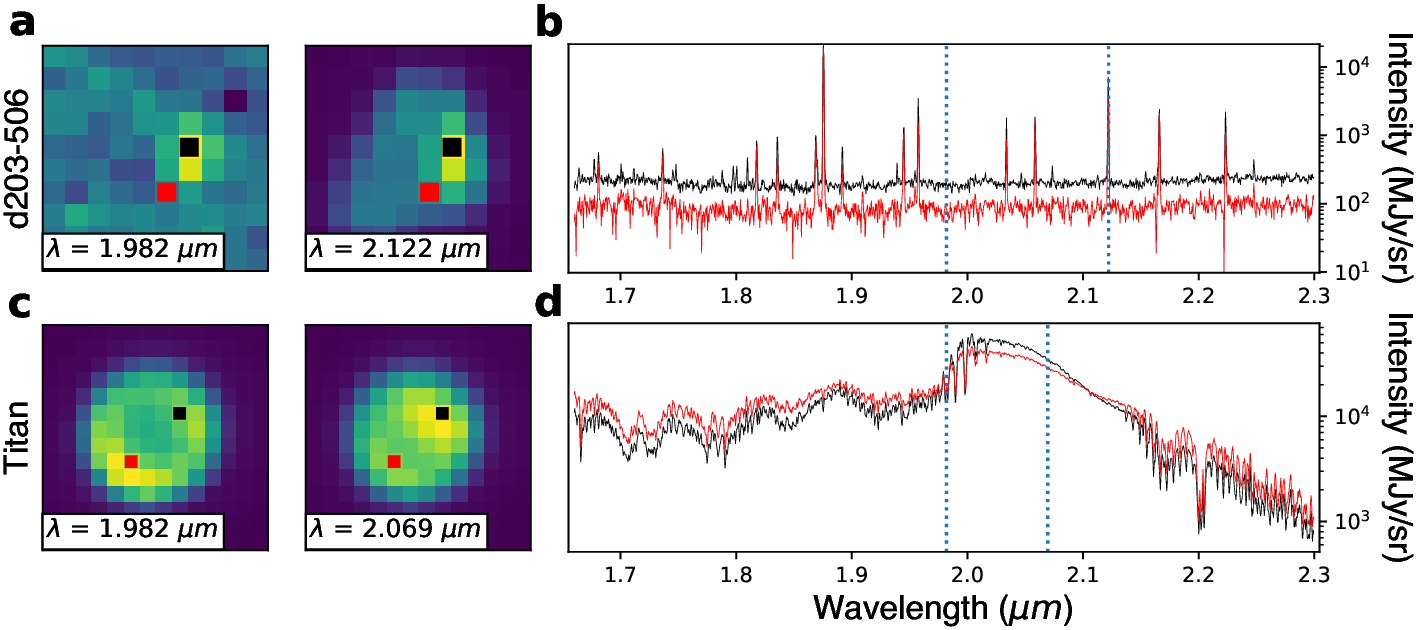}
\centering
\caption{Selected JWST NIRSpec IFU F170LP filter and G235H disperser data. NIRSpec observations of the d203-506 protoplanetary disk at 1.982 and 2.122~$\mathrm{\mu m}$ (\textbf{a}). Spectra from those observations (\textbf{b}) at the positions (red and black points) shown in the images. Titan observed by NIRSpec (\textbf{c}) at 1.982 and 2.069~$\mathrm{\mu m}$. Spectra from those observations (\textbf{d}) at the positions (red and black points) shown in the images. In the plot of the spectra, the two vertical dotted lines indicate the wavelengths of the images. Original data from the MAST database have been rotated, aligned, and cropped (we refer to this as co-registration) as described in Appendix \ref{sec:prepro}.}
\label{fig:NIRSpec}
\end{figure*}

\section{Application to JWST data}\label{sec:results} \addcontentsline{toc}{section}{Results}

Our goal is to fuse pairs of input NIRSpec and NIRCam data that cover the same spectral range and field of view (FoV), a practical context herein referred to as \emph{symmetric} fusion. Fusing real data presents several key challenges: defining the feasible range for its application, finding corresponding data, and ensuring proper pre-processing, such as  alignment and cross-calibration. The methodology to meet these challenges is described in Appendix \ref{sec:prepro}. The proposed so-called Symmetric Fusion (SyFu) algorithm combines the pre-processing and fusion steps. In what follows, we present its application to JWST observations whose specifications are detailed in Appendix \ref{sec:range}.

\subsection{JWST data}\label{subsec:dataset} \addcontentsline{toc}{subsection}{JWST data}

In principle, symmetric fusion of JWST data can be performed across six distinct wavelength ranges (see Fig.~\ref{fig:fusion_range} and Table~\ref{tab:fusion_range}). Yet we decided to restrict the available choices to the NIRCam short wavelength channel ($\lambda < 2.35~\mathrm{\mu m}$) since it provides an angular resolution that is two times higher than the long wavelength channel. This choice thus allows the fusion process to reach the largest expected gain in terms of angular resolution. To ensure the largest spectral overlap between the NIRCam short wavelength channel and NIRSpec, the setup relies on the combination of the NIRSpec G235H disperser and F170LP filter  (see Table~\ref{tab:fusion_range}). Five NIRCam filters cover this wavelength range and fully overlap with G235H/F170LP. Namely, the filters are F182M, F187N, F200W, F210M, and F212N (see Fig.~\ref{fig:fusion_range} and Table~\ref{tab:fusion_range}).

In addition, ensuring that the pair of NIRCam and NIRSpec data to be fused cover the same FoV led us to target JWST science programs with concomitant observations by the two instruments. Among the possible completed observations, we chose two different datasets available through the Mikulski Archive for Space Telescopes (MAST) that are part of two Early Release Science (ERS) and Guaranteed Time Observation (GTO) programs. More precisely, they correspond to the protoplanetary disk d203-506 in the Orion Bar observed within the PDRs4All ERS program \citep{berne2022pdrs4all} and to Titan observed within the GTO program entitled `Titan climate, composition and clouds' \citep{nixon2025atmosphere}. These datasets have been elected to meet the aforementioned criteria defining symmetric fusion. The two pairs of NIRCam and NIRSpec JWST datasets are depicted in Fig.~\ref{fig:NIRCam} and \ref{fig:NIRSpec}, and their respective properties are summarised in Table~\ref{tab:data-information}. In the case of d203-506, it is worth noting that only three filters (namely F182M, F187N, and F210M) are available because no observations were planned through the filter F200W, and those provided by the filter F212N were too noisy  \citep{habart2024pdrs4all}. To mitigate the computational burden induced by the fusion algorithm, the two FoVs of interest where the fusion processes have been conducted were carefully limited to the celestial objects of interest. For the d203-506 dataset, we chose a 1'' $\times$ 1'' FoV centred on the disk or jet (see Figures \ref{fig:NIRCam} and \ref{fig:NIRSpec}). For the Titan dataset, we chose a 1.4'' $\times$ 1.4'' FoV covering the satellite (see Figures \ref{fig:NIRCam} and \ref{fig:NIRSpec}). Finally, we emphasise that the MAST archive may contain additional datasets that satisfy the criteria required for symmetric fusion.

\subsection{Spatial and spectral inspections of the fused cubes} \addcontentsline{toc}{subsection}{Spatial and spectral inspections of the fused hyperspectral cubes}
\label{subsec:fusion_real_data}

\begin{figure*}[t]
    \centering
    \includegraphics[width=0.65\linewidth]{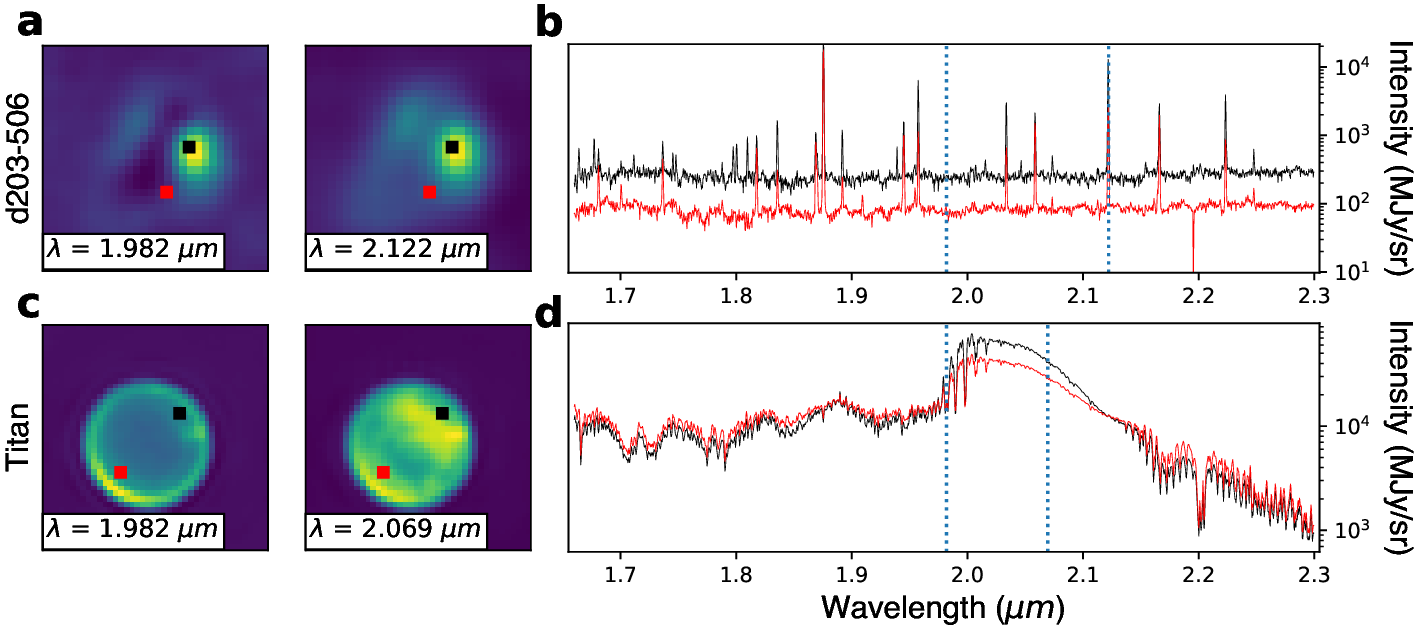}
    \caption{High resolution hyperspectral cubes resulting from JWST data fusion. The d203-506 protoplanetary disk fused hyperspectral cube (\textbf{a}) shown at 1.982 and 2.122~$\mathrm{\mu m}$. Spectra from this cube (\textbf{b}) at the positions (red and black points) shown in the images. Titan fused hyperspectral cube (\textbf{c}) shown at 1.982 and 2.069~$\mathrm{\mu m}$. Two spectra extracted from this cube (\textbf{d}) at the positions (red and black points) shown in the images. In the plot of the spectra, the vertical dotted lines indicate the wavelengths of the images.}
    \label{fig:results}
\end{figure*}

Figure \ref{fig:results} presents spatial and spectral visualisations of the resulting high resolution hyperspectral cubes obtained by the SyFu algorithm, whose parameters and properties are discussed in Appendix \ref{sec:fusion} and reported in Table~\ref{tab:SyFu-parameters-and-information}. The fused hyperspectral cubes achieve an angular resolution close to that of NIRCam while providing spectral information over the 1.66 to 2.3~$\mu$m wavelength range. For the case of d203-506, at 1.98~$\mu$m —corresponding to the Paschen-$\alpha$ line of hydrogen— small-scale structures are well recovered (Fig.~\ref{fig:results}\textbf{a}). Notably, the dark lane caused by the protoplanetary disk silhouetted against the nebular background emission of the Orion Nebula is clearly visible, as is the bright spot associated with the base of a jet \citep{berne2024far}. At 2.12~$\mu$m, which corresponds to the ro-vibrational emission of H$_{2}$ \citep{berne2024far}, the warm wind enshrouding the disk is recovered as well. In the case of Titan, the fused data cube at 1.98~$\mu$m allows for a clear recovery of atmospheric haze and cloud structures (Fig.~\ref{fig:results}\textbf{c}) \citep{nixon2025atmosphere}. At 2.07~$\mu$m, the surface of the satellite is distinctly visible, with the Belet region in the southern hemisphere being discernible \citep{nixon2025atmosphere}. In Fig.~\ref{fig:results}\textbf{b} and \textbf{d}, it appears that the spectra are well recovered over the 1.66 to 2.3~$\mu$m wavelength range. In the case of d203-506, the spectrum extracted from the fused cube (red spectrum in Fig.~\ref{fig:results}\textbf{b}) shows a reduced noise level with respect to the original NIRSpec spectrum (Fig.~\ref{fig:NIRSpec}\textbf{b}). This noise reduction effect is less tractable on Titan data (Fig.~\ref{fig:results}\textbf{d}) because the original NIRSpec cubes (Fig.~\ref{fig:NIRSpec}\textbf{d}) have a higher signal-to-noise ratio than the d203-506 data. Complementary experimental results, in particular assessing the consistency of the fused hyperspectral cubes, are reported in Appendix \ref{sec:consistency}. For reproducibility, Appendix \ref{app:code} contains detailed information supporting the presented results.

\section{Discussion}\label{sec:discussion} \addcontentsline{toc}{section}{Discussion}

The successful data fusion on real astronomical targets presented here opens up new possibilities for analysing astronomical data and designing future observations with JWST and beyond. In the case of JWST, the fused cubes achieve an angular resolution nearly three times higher than the native resolution of NIRSpec. This improvement could have significant implications across several fields of astronomy, as it may, for instance, enable spatially resolving the thousands of emission lines present in the hundreds of irradiated protoplanetary disks in Orion (such as d203-506, presented in Fig.~\ref{fig:results}), which would offer new insights into their physical structure. These targets have recently garnered considerable interest \citep{allen2025past, schroetter2025miri} and have already been observed with NIRCam \citep{berne2022pdrs4all, mccaughrean2023jwst}. Obtaining complementary NIRSpec IFU data would allow a fusion with NIRCam data and ultimately yield unique high-resolution datasets that could greatly enhance our understanding of planetary system formation, as the fusion would allow us to reach astronomical unit scales (the angular resolution of NIRCam corresponds to about 12~AU in Orion).

There are also many promising opportunities for studying planet-forming disks closer to us and for which NIRCam data are already available (e.g. in Taurus L1527 \citep{villenave2024jwst}). In such cases, the improved angular resolution from fusion could enable the identification of spatial structures within disks in specific emission lines.

Nearby galaxies have also been observed with JWST, for instance, as part of the PHANGS program with NIRCam \citep{lee2023phangs}. Targeted NIRSpec observations of specific regions could enable data fusion to study star-forming regions within these galaxies at unprecedented spatial scales and levels of detail. This could pave the way for fusion experiments on galaxies at greater distances. For instance, fusing NIRSpec and NIRCam observations of deep fields could help resolve individual star-forming regions in high-redshift galaxies \citep{Claeyssens_2023}, providing new tools to understand the formation of stellar clusters at cosmic dawn.

A crucial requirement for accomplishing such a fusion is the availability, on the same telescope, of both an imager and spectrograph, as is the case with NIRCam and NIRSpec on JWST. While not unique to JWST, this configuration is still relatively rare. For example, the X-IFU instrument on the future Athena mission could be combined with its imager \citep{lascar2025best}. We advocate for promoting coupled observation modes in future instruments alongside the development of dedicated data fusion algorithms integrated into their pipelines. This practice is already standard in the context of Earth observation.

Although the successful data fusion presented here shows great promise, several improvements are needed to unlock its full potential. One limitation of the current SyFu algorithm is that both the imager and the IFU must observe the same FoV with complete spatial and spectral overlap (i.e. symmetric fusion). A way to expand the applicability of NIRCam–NIRSpec fusion would be to develop non-symmetric fusion methods. This would, for example, enable fusion across the entire NIRSpec wavelength range (0.6 to 5~$\mathrm{\mu m}$), despite the three wavelength gaps in the NIRCam data (Fig.~\ref{fig:fusion_range}).

In this work, we also employed a classical regularisation based on the Sobolev norm. By design, this regularisation smooths the fused hyperspectral cube, which may lead to the loss of fine texture details present in the input image. A more advanced approach would involve weighting the Sobolev regularisation pixel-wise, for instance, by exploiting the input high-resolution image to identify spatial contours that should be preserved and then reducing the smoothing in those regions of the fused cube \citep{guilloteau2022informed}. Other alternatives include informing the regularisation using deep generative models, such as patch normalising flows \citep{altekruger2023patchnr}, to extract local structure from the high-resolution input image and improve the fusion process.

\bibliographystyle{aa}
\bibliography{strings_all_ref, mybiblio}

\begin{appendix}

\section{Data availability}
\label{app:code}
Our fusion results will be shared upon reasonable demand. The original NIRCam and NIRSpec data are available on the MAST archive: \href{https://mast.stsci.edu/portal/Mashup/Clients/Mast/Portal.html}{https://mast.stsci.edu/portal/Mashup/Clients/Mast/Portal.html} (proposal 1288 and 1251). NIRCam throughputs are accessible on the STScI website: \href{https://jwst-docs.stsci.edu/jwst-near-infrared-camera/nircam-instrumentation/nircam-filters}{https://jwst-docs.stsci.edu/jwst-near-infrared-camera/nircam-instrumentation/nircam-filters}.

The SyFu algorithm and the code producing the figures of this paper are available at : \href{https://github.com/L4Marquis/SyFu}{https://github.com/L4Marquis/SyFu}.

\section{Data pre-processing}\label{sec:prepro}
\addcontentsline{toc}{subsection}{Data pre-processing}

The SyFu algorithm includes three pre-fusion steps to compute the forward operators, to co-register the pairs of input data and to cross-calibrate the operators. These stages are detailed in the following sections. To lighten the notations, the three-dimensional data cube was unfolded into matrices whose rows (resp. columns) are associated with the spectral (resp. spatial) dimension.

\subsection*{JWST pipeline processing}

The datasets for d203-506 and Titan are extracted from the MAST archive, hence they already underwent a complete, up to stage 3, JWST science calibration pipeline processing \citep{JWSTpipeline}. This pipeline converts raw observations acquired with an up-the-ramp strategy into science-ready data with common astrophysical units (e.g. MJy/sr). It computes the coordinates, corrects bad pixels and cuts off most of the instrumental noise and cosmic rays present in the data.

\subsection*{Computing the forward operators}\label{subsec:fusion_details}
The two data-fitting term defining the objective function underlying the optimisation problem \eqref{eq:fusion} are driven by the forward models that relate the fused hyperspectral cube to be recovered and the NIRCam and NIRSpec data. The NIRCam forward model writes
\begin{equation}
\textsf{NIRCam}(X) = L_{\mathrm{m}}W_{\mathrm{m}}(X),
\label{eq:NIRCam_forward_model}
\end{equation}
where $L_\mathrm{m}$ is the matrix representing the NIRCam throughputs and $W_\mathrm{m}(\cdot)$ stands for the operator which performs wavelength-wise spatial convolutions with the NIRCam PSFs. Similarly the NIRSpec forward model is decomposed as
\begin{equation}
\textsf{NIRSpec}(X) = L_{\mathrm{h}}W_{\mathrm{h}}(X)S,
\label{eq:NIRSpec_forward_model}
\end{equation}
where $L_\mathrm{h}$ and $W_\mathrm{h}(\cdot)$ denote the NIRSpec throughputs and convolutions by the PSF. The matrix $S$ is a regular sub-sampling matrix that models the difference in resolutions between the NIRCam and NIRSpec data.

\subsection*{Co-registration}\label{subsec:align}

The user defines the rectangular fusion FoV by specifying a central point along with its height and width. This definition only requires a full overlap between NIRCam and NIRSpec data, which characterises a symmetric fusion. Within this fusion FoV, the NIRCam images and NIRSpec cubes must be co-registered, which involves both having a common rotation with respect to the North and aligned pixels.

The NIRSpec cubes are resampled in the JWST pipeline so that the pixels are aligned with North. Since the NIRSpec aperture is not aligned with North at the moment of observation, this implies that the FoV is rotated. To obtain a rectangular aperture aligned with North we rotate the NIRSpec cube with the appropriate angle. We then apply a rotation to the NIRCam image, so that it has the same rotation with respect to the North as the NIRSpec cube.
The next step consists in aligning the NIRCam images and NIRSpec cube. To do so, two cases need to be considered: non-moving (distant) objects and solar system objects. 
In the first case of non-moving objects, applicable to d203-506, we rely on the coordinates to align the NIRCam images and the NIRSpec cube. Finally, the NIRCam images are resampled to a pixel size that is one-third of NIRSpec (0.033''). This division by a factor of three is chosen since it is the closest integer to the pixel scale ratio between NIRSpec and NIRCam short wavelength channel (3.2). Having an integer value provides significant simplifications and speeding-up of the algorithmic implementation \citep{wei2015fast}.
In the second case of moving targets, applicable to Titan, the translations to be operated for the image alignment cannot be deduced from the pointing information. The NIRCam images and NIRSpec cubes are resampled to a common pixel size of one-third of the NIRSpec pixel size (0.033''). Then, using the phase cross-correlation technique \citep{guizar2008crosscorrelation, scikit-image} we compensate for the translations. We note that all mentioned rotation and re-sampling steps were conducted with bicubic interpolations.

\subsection*{Cross-calibration}\label{subsec:cross-calibration}

The SyFu algorithm requires a precise calibration of both NIRCam and NIRSpec forward operators to ensure the spectral consistency of the fused hyperspectral cube. First, the raw PSFs produced by the WebbPSF simulation tool are normalised such that their spreading patterns sum to one at each wavelength. The NIRSpec throughput is already compensated during the flat field correction in stage 2 of the JWST pipeline \citep{JWSTpipelinestage2}, so $L_{\mathrm{h}}$ is an identity matrix. Then, the NIRCam throughputs $L_{\mathrm{m}}$ are interpolated over the NIRSpec wavelengths and cross-calibrated as explained below.

The conventional JWST pipeline procedure to calibrate the throughput $L_{\mathrm{m},f}$ of the $f$-{th} NIRCam filter consists in dividing this throughput by the sum of its coefficients, denoted by $\|L_{\mathrm{m},f}\|_1$ \citep{gordon2022calibration}. However, since the JWST pipeline does not perform cross-calibration, this procedure results in an intensity mismatch between NIRCam data $Y_{\mathrm{m}}$ and NIRSpec data $Y_{\mathrm{h}}$ \citep{chown2024pdrs4all}. To mitigate this mismatch, we cross-calibrate the NIRCam throughputs with respect to the input NIRSpec data $Y_{\mathrm{h}}$ to ensure that
\begin{equation}
    \overline{Y_{\mathrm{m},f}} = L_{\mathrm{m},f} \overline{Y_{\mathrm{h}}},
\label{eq:cross-calibration}
\end{equation}
for every filter $f$, where $Y_{\mathrm{m},f}$ denotes the image of the $f$-{th} NIRCam filter and $\overline{\, \cdot \,}$ denotes the spatial averaging operator. This is done by subsequently multiplying $L_{\mathrm{m},f}$ by the factor
\begin{equation} 
    \|L_{\mathrm{m},f}\|_1 \times \frac{\overline{Y_{\mathrm{m},f}}}{L_{\mathrm{m},f} \overline{Y_{\mathrm{h}}}} ,
\label{eq:ratio}
\end{equation}
where the multiplication by $\|L_{\mathrm{m},f}\|_1$ compensates the conventional calibration procedure and the multiplication by the ratio ensures the identity \eqref{eq:cross-calibration} holds. The resulting cross-calibration factors, close to 1, are reported for each NIRCam filter in Table~\ref{tab:SyFu-parameters-and-information}.

\section{Fusion as a regularised inverse problem}\label{sec:fusion}
\addcontentsline{toc}{subsection}{Fusion by solving an inverse problem with regularisation}
Once the NIRCam and NIRSpec datasets have been co-registered and cross-calibrated, they follow a fusion process by solving the minimisation problem \eqref{eq:fusion}. The strategy adopted to solve this problem is detailed hereafter.

\subsection*{Regularisations}\label{subsec:regul}
The SyFu algorithm performs symmetric fusion by solving a spectrally and spatially regularised inverse problem. Instead of considering a complex joint spatial-spectral regularisation, the regulariser $R(\cdot) = R_{\mathrm{spec}}(\cdot) + R_{\mathrm{spat}}(\cdot)$ in \eqref{eq:fusion} is split into two terms associated with the spectral regularisation and the spatial regularisation, respectively. 

Regarding the spectral regularisation $R_{\mathrm{spec}}(\cdot)$, it is implicitly defined by an \emph{hard} constraint that imposes that the solution $X$ of the minimisation problem lives in the same spectral subspace as the NIRSpec data cube. To identify this subspace we conduct a principal component analysis (PCA) of the NIRSpec data. This procedure involves centring the NIRSpec data by removing the NIRSpec mean spectrum $\overline{Y_{\mathrm{h}}}$ from each spectra of $Y_{\mathrm{h}}$. A truncated singular value decomposition then provides a matrix $V$ whose columns contain $k$ elementary spectra which span the spectral signal subspace. Then the fused hyperspectral cube is decomposed as
\begin{equation}
    X = VZ +  {\Upsilon_{\mathrm{h}}},
\label{eq:PCA}
\end{equation}
where $Z$ is the representation of the solution in the subspace and ${\Upsilon_{\mathrm{h}}} = \left[\overline{Y_{\mathrm{h}}},\ldots, \overline{Y_{\mathrm{h}}}\right]$ is a matrix whose columns are the mean spectrum $\overline{Y_{\mathrm{h}}}$ obtained by spatially averaging the NIRSpec input data. Beyond promoting spectrally consistent solutions, another key advantage of adopting the hardcore regularisation \eqref{eq:PCA} is that it allows the initial regularised least squares problem \eqref{eq:fusion} to be reformulated with respect to the representation $Z$ of the solution in the subspace, significantly decreasing the computational complexity of the optimisation procedure.

Regarding the spatial regularisation, we ensure the fused hyperspectral cube to be spatially smooth by resorting to a Sobolev regularisation $R_{\mathrm{spat}}(\cdot)$ which penalises the gradient of the representation of the fused hyperspectral cube
\begin{equation*}
    R_{\mathrm{spat}}(Z) = \mu \left\|ZD\right\|_{\mathrm{F}}^2,
\label{eq:Sobolev}
\end{equation*}
where  $D$ is the matrix standing for the first-order finite difference operator and $\|\cdot\|_{\mathrm{F}}^2$ denotes the squared Frobenius norm, that is the sum of the squared coefficients of the matrix, the parameter $\mu$ adjusts the weight of the regularisation. Since the columns of $V$ are orthogonal, this choice amounts to defining the spatial regularisation directly on the fused hyperspectral cube $X$ \citep{guilloteau2020hyperspectral}.

\subsection*{Numerical resolution}\label{subsec:resolution}
Given the aforementioned forward models and regularisations, the minimisation problem in \eqref{eq:fusion} is rewritten as
\begin{multline}
            \min_Z \|Y_{\mathrm{m}} - L_{\mathrm{m}}W_{\mathrm{m}}(VZ + {\Upsilon_{\mathrm{h}}})\|_{\mathrm{F}}^2  \\ 
        +  \gamma \|Y_{\mathrm{h}} - L_{\mathrm{h}}W_{\mathrm{h}}(VZ + {\Upsilon_{\mathrm{h}}})S\|_{\mathrm{F}}^2 + \mu \left\|ZD\right\|_{\mathrm{F}}^2.
\label{eq:fusion_Z}
\end{multline}
In particular, thanks to the linearity of the forward model, the NIRCam data fidelity term can be decomposed as
\begin{equation*} 
    \|Y_{\mathrm{m}} - L_{\mathrm{m}}W_{\mathrm{m}}(VZ + {\Upsilon_{\mathrm{h}}})\|_{\mathrm{F}}^2 = \|Y_{\mathrm{m}} - L_{\mathrm{m}}W_{\mathrm{m}}({\Upsilon_{\mathrm{h}}}) - L_{\mathrm{m}}W_{\mathrm{m}}(VZ)\|_{\mathrm{F}}^2.
\end{equation*}
It is worth noting that, thanks to the cross-calibration and the unit normalisation of the PSFs previously discussed, the identity \eqref{eq:cross-calibration} ensures that $Y_{\mathrm{m}} - L_{\mathrm{m}}W_{\mathrm{m}}(\Upsilon_{\mathrm{h}})$ have a near zero mean. 

The resulting problem \eqref{eq:fusion_Z} is smooth and convex and can be solved by any differentiable convex optimisation algorithm. The algorithmic procedure to solve the regularised inverse problem is composed of three successive steps: the computation of the PSF, the computation of a sparse linear system, and finally its resolution. 

Computing the PSFs with WebbPSF takes several hours. They are computed once and then stored to avoid unnecessary re-computation. This can be done for each of the four NIRSpec filter and disperser combinations. To further lighten the computational burden, the associated large convolutional operators $W_{\mathrm{m}}(\cdot)$ and $W_{\mathrm{h}}(\cdot)$ are converted into term-wise multiplications in the Fourier domain. To do so, the associated PSF, as well as the NIRCam and NIRSpec data, are symmetrically padded and evaluated in the Fourier domain using a two-dimensional fast Fourier transform. This formulation enforces periodic boundary conditions, which may result into unwanted boundary effects, for example, when bright sources are present near the edges of the FoV. Denoting by $h$ and $w$ the spatial height and width of the NIRSpec cube $Y_{\mathrm{h}}$, the size of the padded NIRSpec cube and PSF is $3h\times 3w$, while it is $9h\times 9w$ for NIRCam images and PSF. This choice preserves the cross-calibration, ensuring that both the original and padded data and PSF have the same mean.

Then, all quantities defined in \eqref{eq:fusion_Z} are vectorised adopting a sparse matrix representation. By setting the gradient of the resulting objective function to zero, the linear system to be solved is characterised by a large yet structured and highly sparse matrix that can be conveniently represented using the compressed sparse row format (\texttt{CSR}) of the Python \texttt{SciPy} library. Thanks to the highly sparse structure of the linear problem, this vectorisation yields a computational complexity of $\mathcal{O}({p_{\mathrm{m}}k^2})$ for each iteration of the subsequent iterative algorithm, where $p_{\mathrm{m}}$ stands for the number of NIRCam pixels \citep{guilloteau2020hyperspectral}. The matrices associated with this sparse linear system, whose expressions are detailed in \citep{guilloteau2020hyperspectral}, are computed once and stored for subsequent use. This computation takes less than one minute on a laptop when using a reasonably small subspace (here $k=3$, see Table~\ref{tab:SyFu-parameters-and-information}).

Finally, the linear system is solved using the conjugate gradient descent implemented in the \texttt{scipy.sparse.linalg} library, which eases and fastens the tuning of the parameters $\mu$ and $\gamma$ (see Table~\ref{tab:SyFu-parameters-and-information}).  By initialising the algorithm with an interpolated counterpart of the NIRSpec data $Y_{\mathrm{h}}$, 1000 iterations of the algorithm are required to obtain satisfactory results (see also the code publicly available, as detailed in Appendix \ref{app:code}).

\subsection*{Tuning of the algorithmic parameters}\label{subsec:parameters-inverse-problem}

\begin{table}[t]
    \centering\small
    \caption{Parameters and properties of the SyFu algorithm.}
    \begin{tabular}{p{3.4cm}p{2cm}p{2.4cm}}
        \hline
        Parameter and properties & d203-506 & Titan\\
        \hline
        Fusion FoV & 1" x 1" & 1.4" x 1.4" \\
        \hline
        Fusion wave. range ($\mathrm{\mu m}$) & 1.66 - 2.30 & 1.66 - 2.30 \\
        \hline
        Type of scene & Unmoving object & Moving object \\
        \hline
        Cross-calibration coeff. & 1.14, 1.07, 1.13 & 1.06, 1.00, 1.03, 1.00, 1.11 \\
        \hline
        Info. kept after PCA (\%) & 91.58 & 99.96 \\
        \hline
        $k$ & 3 & 3 \\
        \hline
        $\mu$ & $10^{-2}$ & $3.5\times10^{-2}$ \\
        \hline
        $\gamma$ & 0.2695 & 0.1513 \\
        \hline
        Computational time of the linear system & 18.10 s & 44.83 s \\
        \hline
        Computational time for 1000 iterations of the algorithm & 0.11 s & 0.23 s \\
        \hline
    \end{tabular}
    \tablefoot{The type of scene entry refers to the method used to align the data in the alignment section. Cross-calibration coefficients are sorted to match their corresponding NIRCam filters in Table~\ref{tab:data-information} (for d203-506, F182M calibration coefficient is 1.14).}
    \label{tab:SyFu-parameters-and-information}
\end{table}

The resolution of the regularised inverse problems is governed by several parameters, namely the dimension $k$ of the spectral signal subspace, the parameter $\gamma$ which balances the NIRCam and NIRSpec data fitting terms and the parameter $\mu$ which adjusts the weight of the Sobolev regularisation. These parameters can be tuned manually to optimise the results or automatically for non-expert users.\\

\noindent \emph{Manual tuning --} The parameter $k$, which corresponds to the number of principal components to be kept when conducing the PCA, is adjusted to preserve sufficient information in the NIRSpec data. Beyond its impact on the spectral regularisation, it is worth noting that this parameter has also an impact on the dimension of the linear system to be solved. Thus, choosing a small $k$ reduces drastically the computational time required in the iterative minimisation procedure. Regarding the parameter $\gamma$, it has been manually adjusted to reach a trade-off between the two data fitting terms. Finally, the weight of the regularisation $\mu$ is adjusted by performing a grid search with a few values between 1 and $10^{-4}$ and choosing the best one through visual inspection. Table~\ref{tab:SyFu-parameters-and-information} summarises the values of these parameters manually adjusted for the two datasets.\\

\noindent \emph{Automatic tuning --} Techniques commonly adopted in image processing can be devised to automatically guide the selection of the hyperparameters $k$, $\gamma$ and $\mu$. One strategy consists in first estimating the noise levels ${\sigma}_{\mathrm{m}}$ and ${\sigma}_{\mathrm{h}}$ in the NIRCam and NIRSpec data, for example, using the robust estimator proposed by \cite{donoho1995noising}. Then, when implementing the hard spectral regularisation, only the principal components with corresponding eigenvalues higher than $\sigma_\mathrm{h}^2$ should be kept in $V$, explicitly adjusting the number $k$. Finally, a straightforward interpretation of the two other hyperparameters can be offered by adopting an empirical Bayesian formulation of the regularised least-square problem \eqref{eq:fusion_Z}. This interpretation leads to $\gamma={\sigma_\mathrm{m}^2}/{\sigma_\mathrm{h}^2}$ and $\mu = {\sigma_\mathrm{m}^2}/{\alpha^2}$ where $\alpha^2$ is the mean of the squared Frobenius norm of $ZD$, that can be empirically estimated on a crude solution $\hat{Z}$, for example, computed from an interpolated counterpart of the NIRSpec data. This strategy relies on the Gaussian assumption of the residual noises, which is not realistic because of the complex nature of the JWST NIRCam and NIRSpec data. Yet, this simple and fast strategy yields results similar to those reported in the paper.

\section{Data properties and range of symmetric fusion}\label{sec:range} \addcontentsline{toc}{section}{Range of symmetric fusion}

\begin{figure}[h!]
\centering
\includegraphics[width=0.5\textwidth]{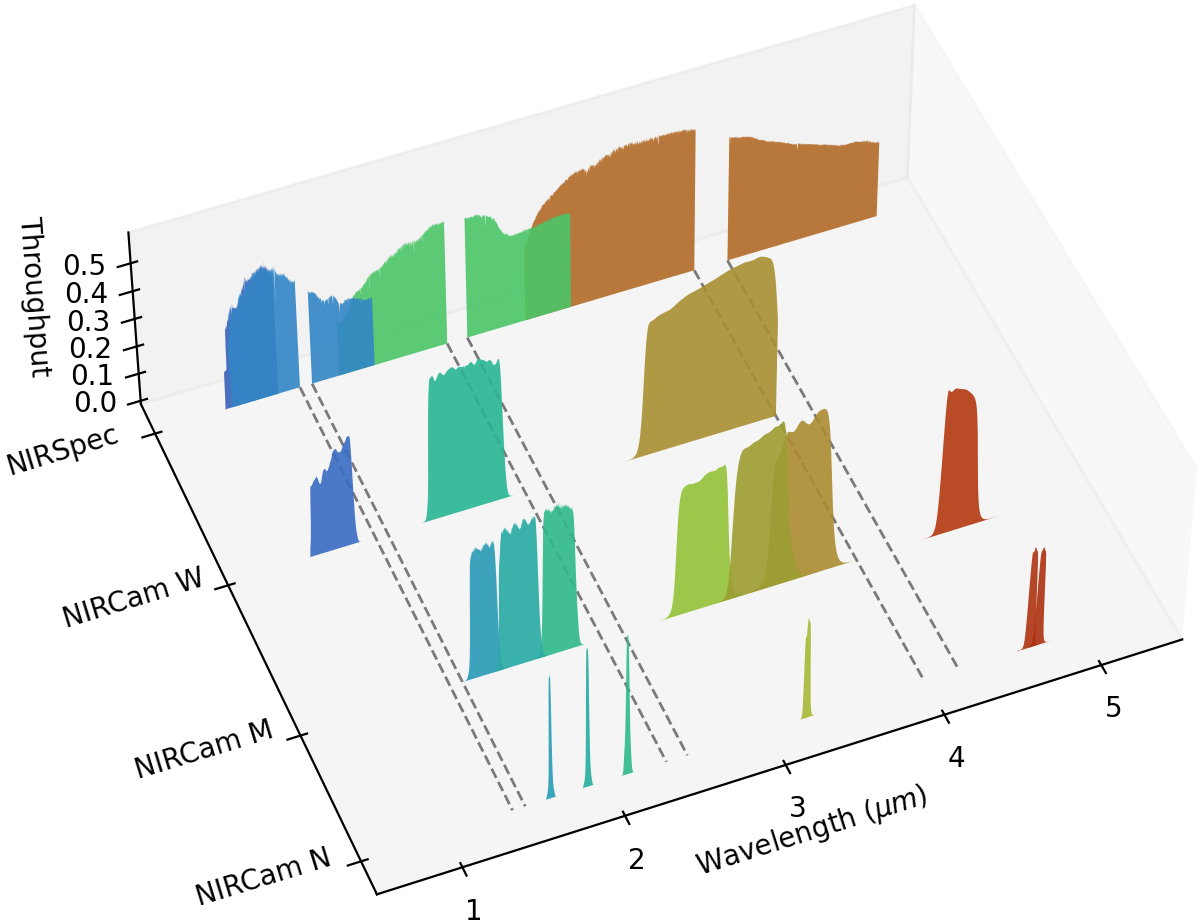}
\caption{NIRCam and NIRSpec throughputs available for a symmetric fusion. The throughputs are sorted on four categories (from top to bottom): NIRSpec high resolution disperser/filter combinations, NIRCam wide (W) filters, NIRCam medium (M) filters and NIRCam narrow (N) filters. The three wavelength gaps (1.40780 - 1.4858~$\mathrm{\mu m}$, 2.36067 - 2.49153~$\mathrm{\mu m}$ and 3.98276 - 4.20323~$\mathrm{\mu m}$) of NIRSpec IFU, where the information is only partial, are limited by the grey dashed lines. NIRCam filters covering those gaps are not displayed.}\label{fig:fusion_range}
\end{figure}

The NIRCam and NIRSpec data extracted from the Barbara Ann Mikulski Archive for Space Telescope (MAST) database are fused on the 3" x 3" FoV of NIRSpec (Table \ref{tab:data-information}). The wavelength ranges on which a symmetric fusion can be performed are limited by:
\begin{enumerate}
    \item the presence of three wavelength gaps in NIRSpec high resolution gratings \citep{STScINIRSpecIFUoverview} (see Fig.~\ref{fig:fusion_range}),
    \item the range of NIRCam filters, whose spectral information must be fully contained in NIRSpec throughputs.
\end{enumerate}
From those constraints we extracted the six different wavelength ranges available for symmetric fusion (see Table \ref{tab:fusion_range}).

\begin{table}[h!]
    \centering\small
    \caption{Main properties of d203-506 and Titan data.}
    \begin{tabular}{p{3.5cm}p{2cm}p{2.2cm}}
        \hline
        Property & d203-506 & Titan\\
        \hline
        Program ID & ERS 1288 & GTO 1251 \\
        \hline
        NIRCam filters & F182M, F187N, F210M & F182M, F187N, F200W, F210M, F212N \\
        \hline
        NIRCam module & B2 & B1 \\
        \hline
        NIRCam FoV & 2.2' x 2.2' & 5.5" x 5.5" \\
        \hline
        NIRSpec filter/disperser & F170LP/G235H & F170LP/G235H \\
        \hline
        NIRSpec FoV & 3" x 3" & 3" x 3" \\
        \hline
        NIRSpec wave. range ($\mathrm{\mu m}$) & 1.66 - 2.36 & 1.66 - 2.36 \\
        \hline
        Pipeline version & 1.16.0 & 1.14.0 \\
        \hline
    \end{tabular}
    \tablefoot{The two datasets come from the Barbara Ann Mikulski Archive for Space Telescope (MAST) database.}
    \label{tab:data-information}
\end{table}

\begin{table}[h]
    \centering\small
    \caption{Combination of NIRCam filters and NIRSpec disperser/filter available for symmetric fusion.}
    \begin{tabular}{p{2cm}p{3.4cm}p{2.1cm}}
        \hline
        Wavelength range ($\mathrm{\mu m}$) & NIRCam filters & NIRSpec disperser/filter\\
        \hline
        0.97 - 1.27 & F115W & G140H/F070LP \\ 
        \hline
        0.97 - 1.355 & F115W & G140H/F100LP \\
        \hline
        1.49 - 1.89 & F162M, F164N & G140H/F100LP \\
        \hline
        1.66 - 2.3 & F182M, F187N, F200W, F210M, F212N & G235H/F170LP \\
        \hline
        2.87 - 3.98 & F300M, F323N, F335M, F356W, F360M & G395H/F290LP \\
        \hline
        4.35 - 5.1 & F460M, F466N, F470N & G395H/F290LP \\
        \hline
    \end{tabular}
    \label{tab:fusion_range}
\end{table}

\section{Consistency of the fused hyperspectral cubes}\label{sec:consistency}
\addcontentsline{toc}{subsection}{Consistency of the fused hyperspectral cubes}

\begin{figure*}[t]
\centering
\includegraphics[width=0.85\textwidth]{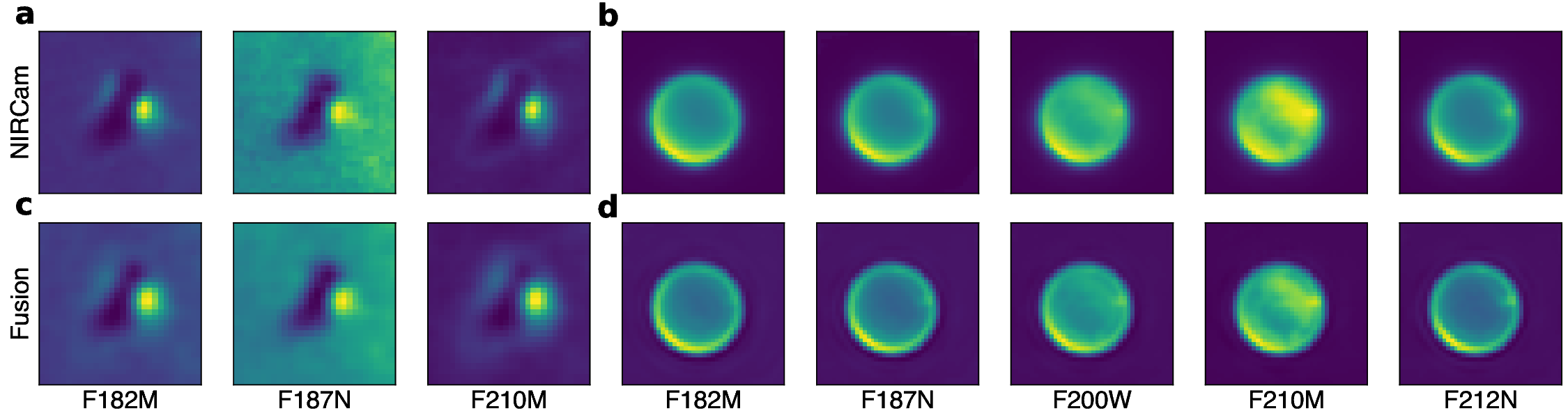}
\caption{Comparison between JWST NIRCam images and images from the fused data cube. 
NIRCam images of the d203-506 protoplanetary disk \textbf{a} and Titan \textbf{b}. 
Images extracted from the fused cube by integrating over the NIRCam throughputs for d203-506 \textbf{c} and Titan \textbf{d}.}\label{fig:val_NIRCam}
\end{figure*}

\begin{figure*}[t]
\centering
\includegraphics[width=0.8\textwidth]{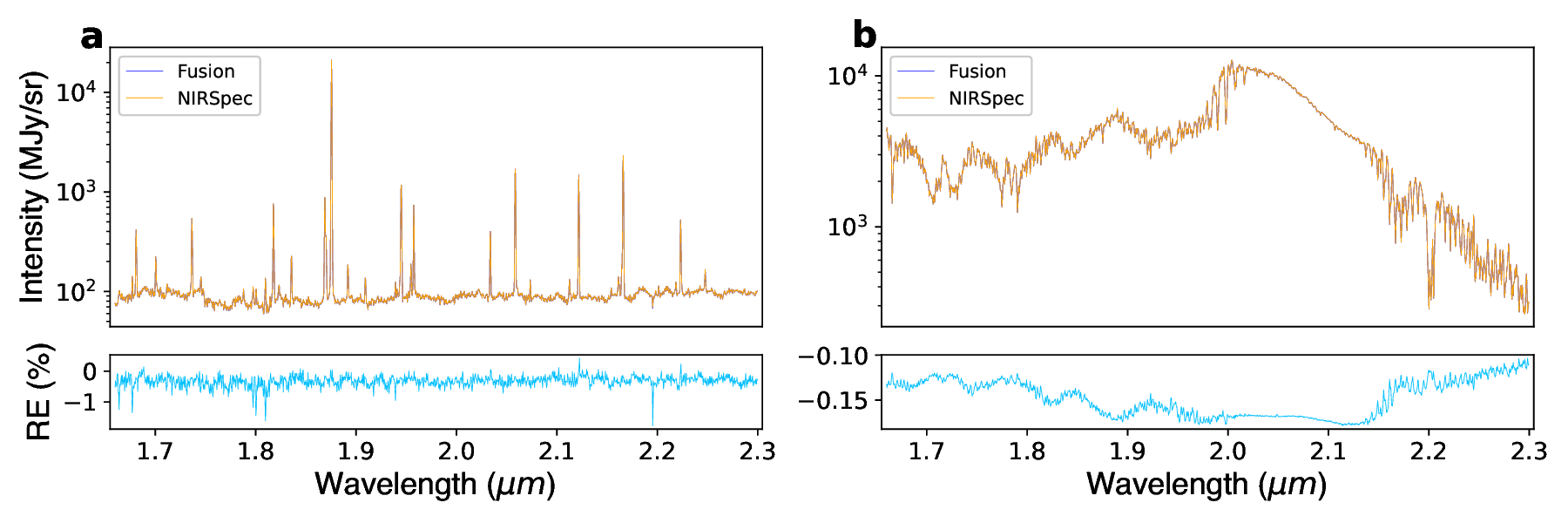}
\caption{Comparison between the JWST NIRSpec and the fused data cube averaged spectra. \textbf{a}, d203-506 protoplanetary disk averaged NIRSpec spectrum (orange) and fused hyperspectral cube averaged spectra (dark blue). Relative error (in percent) between them is shown below, in blue. \textbf{b} : Same as \textbf{a}, but for Titan.}\label{fig:val_NIRSpec}
\end{figure*}

\subsection{Measures of quality}

\begin{figure*}[t!]
\centering
\includegraphics[width=0.7\textwidth]{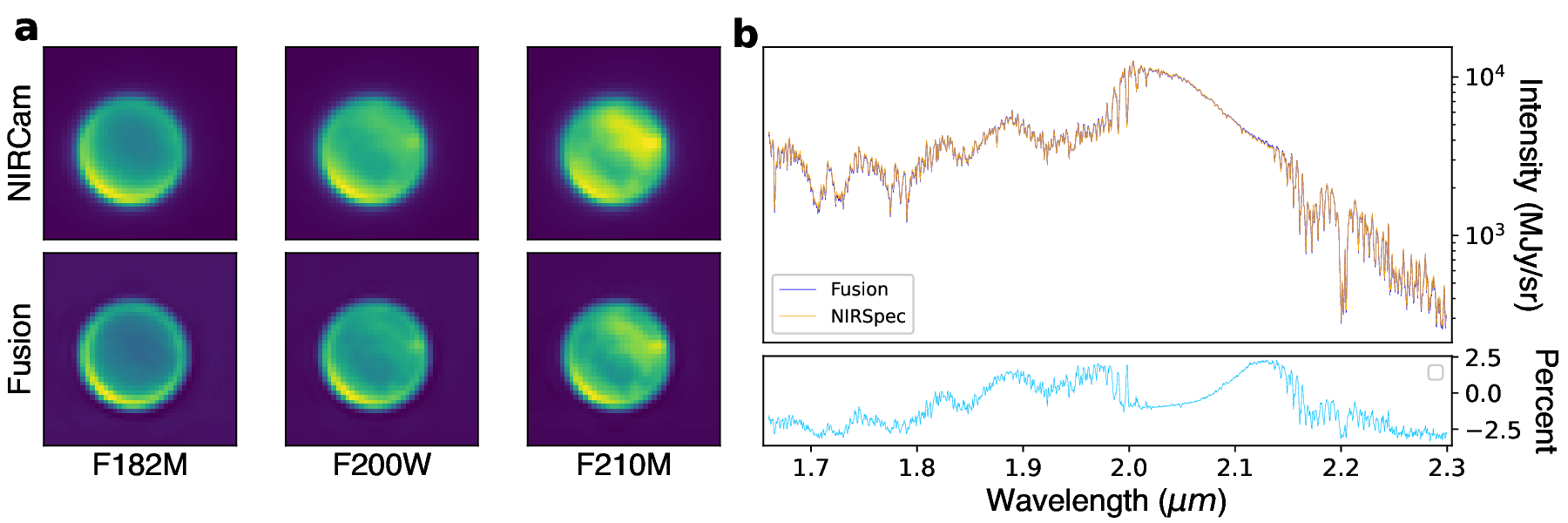}
\caption{Comparison between the JWST NIRCam and NIRSpec Titan data and the hyperspectral cube resulting from their fusion. \textbf{a}: Images of Titan observed by NIRCam filters F182M, F200W and F210M (top) and the Titan fused hyperspectral cube integrated over the corresponding throughputs (bottom). \textbf{b}, NIRSpec averaged spectrum (orange), the fused hyperspectral cube average spectrum (dark blue). Relative error (in percent) between them is shown below, in blue.}\label{fig:validation_marginals}
\end{figure*}

\begin{figure}[t]
\centering
\includegraphics[width=0.25\textwidth]{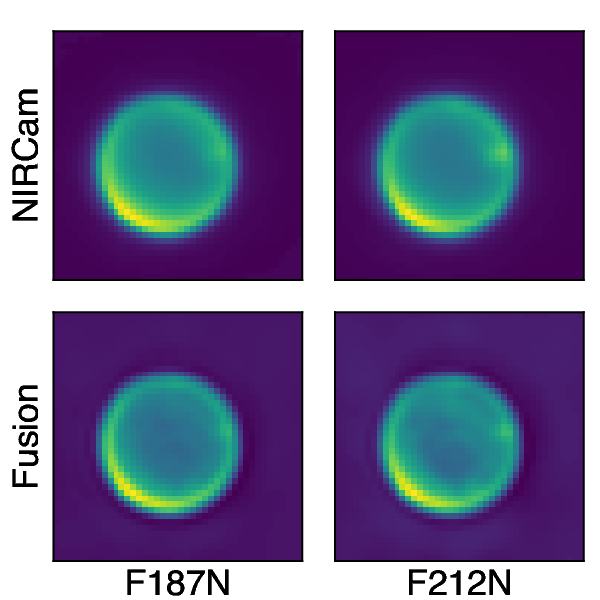}
\caption{The two NIRCam Titan images excluded from data fusion (top) and the fused hyperspectral cube integrated over their respective throughputs (bottom).}\label{fig:unseen_filters}
\end{figure}

The fused hyperspectral cubes (Fig.~\ref{fig:results}) are spectrally degraded using the NIRCam forward model in equation \eqref{eq:NIRCam_forward_model}, producing $F$ images denoted by $X_{\mathrm{m},f}$ for $f \in [\![1;F]\!]$. We then compare these reconstructed images with the NIRCam measurements $Y_{\mathrm{m},f}$, by means of two metrics widely used in image processing. The first one, denoted by peak signal-to-noise ratio (PS/N), computes a normalised relative error between images $X$ and $Y$:
\begin{equation}
    \mathrm{PS/N}(Y,X) = 10 \mathrm{log}_{\rm 10}\frac{\max{(Y)}^2}{\mathrm{MSE}(Y, X)},
\end{equation}
where $\mathrm{MSE}(\cdot,\cdot)$ denotes the mean squared error. The better the reconstruction, the higher the $\mathrm{PS/N}$.
In addition, we compute the structural similarity index~\citep{wang2004image}, which is a local measure of similarity between two image patches $X$ and $Y$, designed to match human perception:
\begin{equation}
    \mathrm{SSIM}(Y, X) = \frac{(2 \overline{Y} \times \overline{X}+c_1)(2\sigma(Y, X)+c_2))}{(\overline{Y}^2 \times \overline{X}^2+c_1)(\sigma(Y)^2+\sigma(X)^2+c_2)},
\end{equation}
where $\sigma(\cdot, \cdot)$ is the covariance, $\sigma(\cdot)$ is the standard deviation, $c_1 = (0.01 (\max(Y)- \min(Y))^2$ and $c_2 = (0.03 (\max(Y)- \min(Y))^2$ are two small variables stabilising the division \citep{wang2004image}. The SSIM is computed locally on  patches of size $7\times 7$, and then averaged over all patches in the image. The better the reconstruction, the higher the $\mathrm{SSIM}$ with a maximum of $1$.
The numerical values are reported in Table~\ref{tab:PS/N}.

\begin{table}[h!]
    \centering\small
    \caption{Measures of the fidelity of the fused hyperspectral cubes with NIRCam images.}
    \begin{tabular}{p{2cm}p{1.5cm}p{2cm}p{1.5cm}}
        \hline
        dataset & Filter & PS/N (dB) & SSIM \\
        \hline
        \multirow{3}{*}{d203-506} & F182M & 30.14 & 0.907 \\
         & F187N & 40.14 & 0.966 \\
         & F210M & 29.24 & 0.903 \\
        \hline
        \multirow{5}{*}{Titan} & F182M & 39.95 & 0.995 \\
         & F187N & 40.33 & 0.988 \\
         & F200W & 42.91 & 0.997 \\
         & F210M & 44.40 & 0.997 \\
         & F212N & 46.17 & 0.998 \\
        \hline
    \end{tabular}
    \label{tab:PS/N}
\end{table}

\subsection{Consistency with the input}

The consistency of the SyFu method can be evaluated by conducting a sanity check which assesses the spatial and spectral consistency of marginal products derived from the fused hyperspectral cubes with respect to the input NIRCam and NIRSpec data.  
More precisely, we first integrate the fused hyperspectral cubes over the throughputs of the NIRCam filters used for fusion. These images extracted from the fused cube are compared to the corresponding input NIRCam images in Fig.~\ref{fig:val_NIRCam}. The extracted images match very well the NIRCam images. This is also supported by the results on Table \ref{tab:PS/N}, where all PS/N values are $\gtrsim$ 30 dB (resp $\gtrsim$ 40 dB) and all SSIM values are $>$ 0.9 (resp. $\gtrsim$ 0.99) for d203-506 data (resp. Titan data), which is recognised as a good pixel-wise and perceptual similarity in the image processing literature.

Second, we average the fused hyperspectral cubes over the two spatial dimensions. The resulting mean spectra are compared to the spatially averaged NIRSpec spectra in Fig.~\ref{fig:val_NIRSpec}. These two spectra are nearly undistinguishable. To quantify the differences, we compute the relative error (RE) between the fused image $X$ and the NIRSpec input data $Y_{\mathrm{h}}$, defined  as $(\overline{{X}} - \overline{Y_{\mathrm{h}}}) \textfractionsolidus {\overline{Y_{\mathrm{h}}}}$ where $\bar{\cdot}$ stands for the spatial averaging operator. These relative errors associated with the two datasets are depicted in Fig.~\ref{fig:val_NIRSpec} as functions of the wavelength. Over the considered spectral range, they are lower than two percent for d203-506 data and even less than 0.2 percent for Titan data.

Overall, Fig.~\ref{fig:val_NIRCam} and \ref{fig:val_NIRSpec} demonstrate that the fused hyperspectral cubes produced by the proposed SyFu algorithm are highly consistent with both NIRCam and NIRSpec observations. We show that more sophisticated validation tests are also satisfied in the following.

\subsection{Additional validation tests}

To further assess the consistency of the data fusion method, we considered the previously described Titan dataset and rerun the fusion, but excluding NIRCam data for filters F187N and F212N. Figure ~\ref{fig:validation_marginals}\textbf{a, b} shows the marginal products derived from the resulting fused hyperspectral cube (as in Fig.~\ref{fig:val_NIRCam} and \ref{fig:val_NIRSpec}) which demonstrate it pass the sanity check described in the Results section. We further validate the consistency of the fused hyperspectral cube in Fig.~\ref{fig:unseen_filters} where the unused F187N and F212N images are compared with the fused hyperspectral cube integrated over their respective throughputs. From a visual inspection, those images match each other closely, which is supported by respective PS/N values of 34.03 and 28.43 (see definition below). 
To conclude, the SyFu algorithm can generate hyperspectral cubes that are fully consistent even with NIRCam observations excluded from the fusion.

\end{appendix}

\end{document}